\documentclass{iopjournal}

\usepackage{amsmath}
\usepackage{amssymb}
\usepackage{bm}

\newcommand{\mps}{m_{\rm PS}}
\newcommand{\qq}{\langle qq\rangle}
\newcommand{\qbarq}{\langle\bar q q\rangle}
\newcommand{\nq}{\langle n_q\rangle}
\newcommand{\tr}{\mathrm{tr}}

\begin{document}

\articletype{Paper}

\title{Equation of state of QCD(-like) theory using Lattice Monte Carlo simulations}

\author{Etsuko Itou$^{1,2}$}

\affil{$^1$Yukawa Institute for Theoretical Physics, Kyoto University, Kitashirakawa Oiwakecho, Sakyo-ku, Kyoto 606-8502, Japan}
\affil{$^2$Interdisciplinary Theoretical and Mathematical Sciences Program (iTHEMS), RIKEN, Wako, Saitama 351-0198, Japan}

\email{itou@yukawa.kyoto-u.ac.jp}


\begin{abstract}
The equation of state (EoS) of strongly interacting matter at low temperature and high baryon density is a central ingredient in the physics of compact stars, but it is still difficult to determine directly from first-principles QCD calculations because of the sign problem. Two-colour QCD (QC$_2$D) provides a useful and controllable laboratory: for an even number of fundamental flavours the fermion determinant is real and positive, while the theory shares the nonperturbative properties with three-colour QCD at least zero chemical-potential regime. 
In this short review we summarize recent lattice progress on dense QC$_2$D, with emphasis on the phase structure, the emergence of superfluidity, the Bose--Einstein-condensation (BEC) to Bardeen--Cooper--Schrieffer(BCS) crossover, and thermodynamic quantities. 
A particularly interesting outcome is that the sound velocity in the cold dense superfluid regime can exceed the conformal value $c_{\rm s}^2/c^2=1/3$. This behaviour, now seen in the simulations for several QCD-like theories, gives a  stiff strongly interacting matter.
It is expected to offer insight into at least some aspects of the physics realized inside neutron stars.
\end{abstract}

\section{Introduction}

Lattice QCD has become a quantitatively reliable nonperturbative method for the vacuum and finite-temperature regimes of strong interactions. 
The hadron spectrum and the chiral crossover temperature are now known with high precision from simulations with dynamical quarks, and the thermal equation of state (EoS) has also been determined precisely~\cite{Borsanyi:2013bia,HotQCD:2014kol}.
In particular, the thermal EoS provides the energy density and pressure that enter the energy--momentum tensor in the Friedmann equations used to describe the evolution of the Universe.
The situation is much less satisfactory in the cold and dense region. Once a quark chemical potential $\mu$ is introduced, the Euclidean Dirac determinant becomes complex in three-colour QCD, and ordinary importance sampling methods cannot be applied. This is the well-known sign problem; in its generic form, it is known to be NP-hard~\footnote{Here, NP (nondeterministic polynomial time) denotes the complexity class of decision problems for which a proposed answer can be verified in polynomial time. NP-hard means that the problem is at least as hard as any problem in NP.}, and its computational cost is expected to grow rapidly toward the thermodynamic limit~\cite{Troyer:2004ge,Aarts:2015kea,Nagata:2021ugx}.

There are nevertheless QCD-like theories in which dense matter can be studied directly. The most popular examples are QCD with isospin chemical potential and two-colour QCD (QC$_2$D). In the latter theory, the pseudo-reality of the fundamental representation of ${\rm SU}(2)$ makes the fermion determinant real, and for an even number of flavours it becomes positive. Therefore Hybrid Monte Carlo simulations based on the importance sampling method can be performed. QC$_2$D\ is not real-world QCD: its baryons are bosonic diquarks, and the superfluid condensate is a colour singlet. Even so, it is a valuable testing ground because it shares confinement, chiral symmetry breaking and asymptotic freedom with ordinary QCD at least zero chemical potential regime.

Among the quantities that can be studied in dense QCD-like theories, the EoS and the sound velocity are of particular interest. They are also directly connected to current questions in compact-star physics: 
neutron-star observations, through masses and radii relation, constrain the EoS of cold dense matter, and many phenomenological analyses suggest that the speed of sound may become large inside massive neutron stars~\cite{Masuda:2012ed,Baym:2017whm,Altiparmak:2022bke,Brandes:2022nxa}.
On the other hands, a reference value so-called the conformal bound,
\begin{equation}
  c_{\rm s}^2/c^2 = \frac{1}{3},
  \label{eq:conformal}
\end{equation}
has been discussed in some holographic models~\cite{Hohler:2009tv,Cherman:2009tw}.
Thus, the sound velocity has an upper bound, which is the value at an ideal relativistic gas.
Lattice results for QCD and QCD-like theories at least at finite temperatures has not shown any violation of this value (see e.g. Refs.~\cite{Borsanyi:2013bia, HotQCD:2014kol}).
Whether it is a true bound in cold dense QCD remains unknown, and recent results in sign-problem-free QCD-like theories provide a useful arena to address this question.

This review is organized as follows. Section~\ref{sec:def-2colorQCD} summarizes the lattice formulation used in our series of QC$_2$D studies. Section~\ref{sec:phase} reviews the phase structure, including the hadronic phase, the superfluid phase, the BEC--BCS crossover and gluonic observables. Section~\ref{sec:eos} discusses the EoS and the sound velocity. Section~\ref{sec:summary} gives summary and outlook.

%%%%%%%%%%%%%%%%%%%%%%%%%%%%%%%%%%%%%%%%%%%%%%%%%%%%%%%%%%
\section{Two-colour QCD on the lattice}\label{sec:def-2colorQCD}
%%%%%%%%%%%%%%%%%%%%%%%%%%%%%%%%%%%%%%%%%%%%%%%%%%%%%%%%%%
\subsection{QC$_2$D lattice action to study cold and dense regime}
The Lagragian here we consider is given by
\begin{equation}
    \mathcal{L}=-\frac{1}{4}F_{\mu \nu}^a F_{\mu \nu}^a +\bar{\psi} (i \gamma_\mu D_\mu + m )\psi +\mu \bar{\psi}\gamma_0 \psi - j [\bar{\psi}_1 (C \gamma_5) \tau_2 \bar{\psi}_2^{T} -  \psi_2^T (C \gamma_5) \tau_2 \psi_1],\label{eq:cont-action}
\end{equation}
in the continuum limit. We mainly consider two-flavour QC$_2$D in this manuscript.
The first two terms are the same as those in the standard QCD Lagrangian, and the third term is the density term associated with the quark number operator. The term proportional to $j$ is the diquark source term, which explicitly breaks the U(1)$_B$ symmetry. In a finite volume, such an explicit breaking term is introduced to study the spontaneous breaking of U(1)$_B$ through the diquark condensate, and the physical result is obtained by taking the $j \to 0$ limit.

In the lattice simulations, we use the Iwasaki gauge action and two flavours of Wilson fermions~\cite{Iida:2019rah,Iida:2020emi,Iida:2022hyy,Iida:2024irv}. 
The action for quarks is expressed by
\begin{equation}
  S_F=\bar\psi_1\Delta(\mu)\psi_1+\bar\psi_2\Delta(\mu)\psi_2
  -J\bar\psi_1(C\gamma_5)\tau_2\bar\psi_2^T
  +J\psi_2^T(C\gamma_5)\tau_2\psi_1 ,
  \label{eq:source}
\end{equation}
where $\Delta(\mu)$ denotes the Wilson--Dirac operator includes the density term. The diquark source parameter $j$ becomes $J=j\kappa$, where $\kappa$ denotes the hopping parameter coming from the normalization of the Wilson fermion on the lattice.
For ${\rm SU}(2)$ gauge fields, the link variable satisfies $U_\mu^\ast=\tau_2 U_\mu\tau_2$. The Dirac operator then obeys
\begin{equation}
  \Delta(\mu)^\ast=\tau_2(C\gamma_5)\Delta(\mu)(C\gamma_5)^{-1}\tau_2 ,
  \label{eq:pseudoreal}
\end{equation}
and hence $\det\Delta(\mu)$ is real. Furthermore, the fermion action, Eq.~\eqref{eq:source}, takes positive if we consider two degenerate masses for flavours. This is the reason why even-flavor dense QC$_2$D\ can be simulated by standard Monte Carlo methods.

The absence of the sign problem is not the whole story. 
Around the onset scale $\mu\simeq \mps/2$, where $\mps$ denotes the mass of the pseudo-scalar meson (pion) at $\mu=0$, numerical instabilities arise coming from the emergence of the zero eigenvalue of the Wilson-Dirac operator~\cite{Muroya:2000qp, Muroya:2002jj}.
This numerical instability can be controlled by introducing a small diquark source term proportional to $J$. 
Technically, this source lifts the eigenvalues of the Wilson--Dirac operator, thereby stabilizing the simulation.
In Monte Carlo simulations, the fermionic degrees of freedom are integrated out analytically, leaving the action written only in terms of the gauge fields. The presence of the diquark source term requires a corresponding modification of this Gaussian integration. The fermion action is therefore rewritten with an extended matrix, and Rational Hybrid Monte Carlo is used for the two-flavour theory (see detail Ref.~\cite{Iida:2019rah, Iida:2024irv}).

The numerical results reviewed here were obtained at $(\beta,\kappa)=(0.800,0.159)$.
The relative scale setting was obtained from the gradient-flow scale $w_0$ and the pseudocritical temperature at $\mu=0$ \cite{Iida:2020emi}. With the convention $T_c\simeq200$ MeV, the lattice spacing is approximately $a=0.17$ fm, and the pseudoscalar mass is $m_{\rm PS}\simeq738$ MeV. The low-temperature data have been obtained on $16^4$ and $32^4$ lattices, corresponding roughly to $T\simeq80$ MeV and $T\simeq40$ MeV, respectively \cite{Iida:2019rah,Iida:2024irv}.

The role of this scale setting should be kept in mind. Since QC$_2$D\ is not a theory of nature, there is no unique experimental quantity with which to set the absolute scale. The convention above is chosen so that temperatures and chemical potentials can be displayed in familiar units. The more robust quantities are dimensionless ratios such as $T/T_c$, $\mu/\mps$ and $m_{\rm PS}/m_{\rm V}$. In particular, the statement that the superfluid onset occurs at $\mu\simeq \mps/2$ is independent of the arbitrary conversion to MeV.

There are two practical upper limits on the chemical potential. The first is the lattice saturation effect, in which all lattice sites become filled with quarks. The second, usually more restrictive in the parameter range of present simulations, is the appearance of large $a\mu$ artifacts before the saturation plateau is reached. 
In lattice simulations, the lattice spacing $a$ provides the UV cutoff, and dimensionless quantities such as $a\mu$ should remain smaller than unity for reliable continuum-oriented physics. In our simulations, we indeed observed unphysical behaviour already around $a\mu \sim 0.8$, where quantities that should monotonically increase with $\mu$ started to decrease. In this review, we therefore do not consider the parameter region affected by such lattice artifacts from the physics discussion.

%%%%%%%%%%%%%%%%%%%%%%%%%%%%%%%%%%%%%%%%%%%%%%%%%%%%%%%%%
\section{Phase structure}\label{sec:phase}
%%%%%%%%%%%%%%%%%%%%%%%%%%%%%%%%%%%%%%%%%%%%%%%%%%%%%%%%%
\subsection{Observables}
\begin{figure}
\centering
\includegraphics[width=0.88\textwidth]{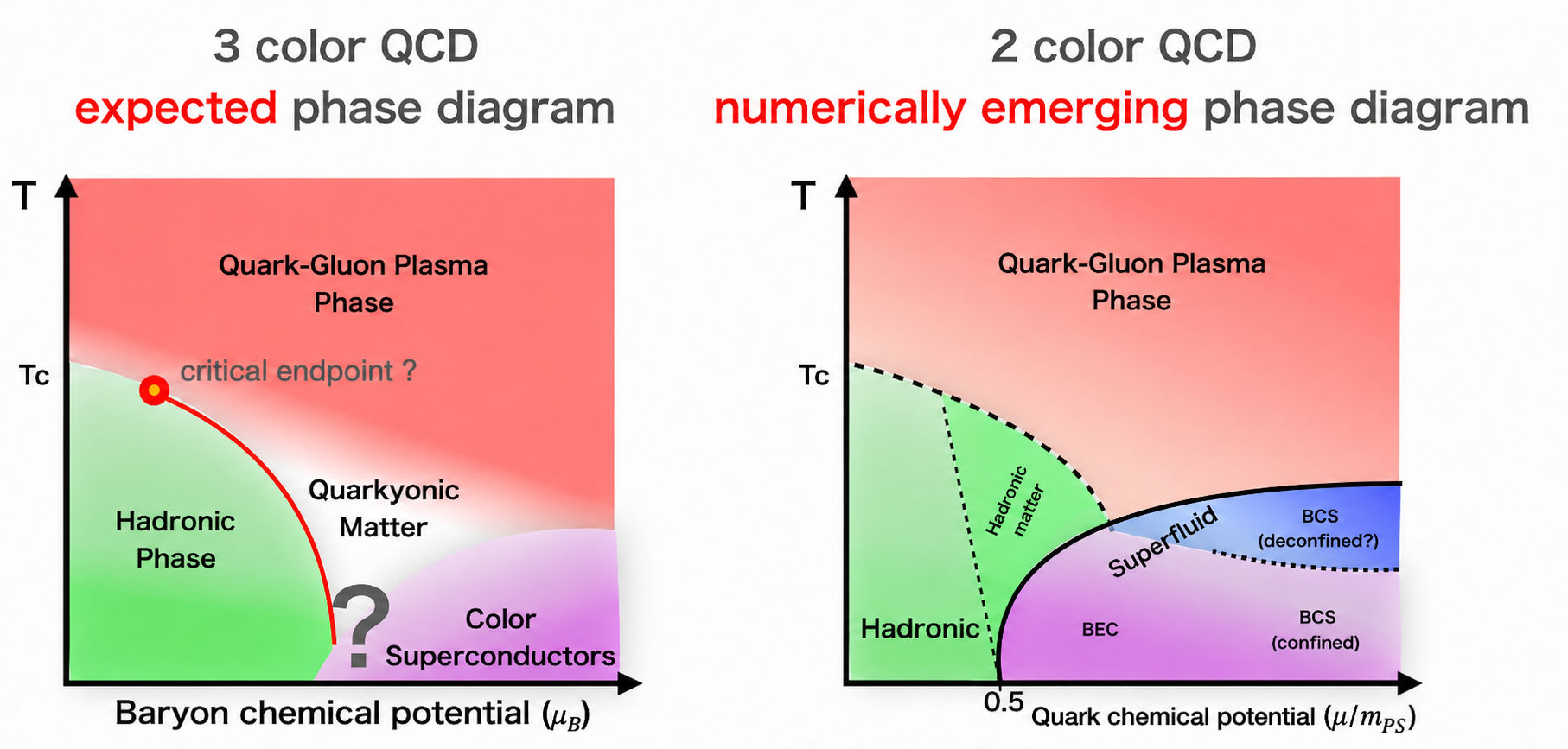}
\caption{Schematic comparison of the phase diagrams of three-colour QCD and two-colour QCD. The three-colour QCD panel is based on the phase-diagram review of Fukushima and Hatsuda \cite{Fukushima:2010bq}; the QC$_2$D\ panel summarizes the lattice picture obtained in Ref.~\cite{Iida:2024irv}.}
\label{fig:phase-diagram}
\end{figure}

Before turning to the EoS, let us first look at the ground-state structure in the region of interest. Figure~\ref{fig:phase-diagram} shows a schematic comparison of the phase diagrams. The left panel shows the commonly expected phase structure of three-colour QCD, where the finite-baryon-density region is difficult to access directly from lattice simulations. The right panel shows the phase structure that has emerged from recent QC$_2$D simulations (See a recent review~\cite{Itou:2025vcy}). In the following, we introduce the physical observables used to distinguish these phases and give their definitions.
%%%%%%%%%%%%%%%%%%%%%%%%%%
%%%%%%%%%%%%%%%%%%%%%%%%%%

The Polyakov loop is used as a diagnostic of confinement,
\begin{equation}
  L=\frac{1}{N_s^3}\sum_{\bm{x}}\tr\prod_\tau U_4(\bm{x},\tau).
  \label{eq:polyakov}
\end{equation}
The diquark condensate is the order parameter for the superfluid transition,
\begin{equation}
  \qq=\frac{\kappa}{2}
  \left\langle
  \bar\psi_1(C\gamma_5)\tau_2\bar\psi_2^T
  -\psi_2^T(C\gamma_5)\tau_2\psi_1
  \right\rangle.
  \label{eq:diquark}
\end{equation}
Also, the (net) quark number density,
\begin{equation}
    a^3 \langle n_q \rangle= \sum_{i} \kappa \langle \bar{\psi}_i (x) (\gamma_4 -\mathbb{I}_4) e^\mu U_{4} (x) \psi_i (x+\hat{4})  + \bar{\psi}_i (x) (\gamma_4 + \mathbb{I}_4) e^{-\mu}U_4^\dag (x-\hat{4} )\psi_i (x-\hat{4}) \rangle,
\end{equation}
represents the difference between quark and antiquark densities and is used for identifying the crossover from the BEC-like regime to the BCS-like regime. 
The chiral condensate and the topological susceptibility are not used to define the phases in our terminology, but they provide important information about the vacuum structure at finite density.

Table~\ref{tab:phases} summarizes the definitions of these phases in terms of the behaviour of the corresponding observables. 
%%%%%%%%%%%%%%%%%%%%%%%%%%
\begin{table}[h]
\caption{Definition of each phase of dense QC$_2$D. The BEC and BCS regimes are continuously connected inside the superfluid phase, but they are distinguished by the quantitative behaviour of the quark number density.}
\centering
\begin{tabular}{l c c c}
\hline
Regime & $\langle |L|\rangle$ &$\qq$ & $\nq$ \\
\hline
Hadronic & zero &zero & zero  \\
Hadronic-matter & zero & zero & nonzero \\
BEC superfluid & & nonzero & nonzero  \\
BCS superfluid & & nonzero & $\simeq n_q^{\rm tree}$ \\
QGP & nonzero &zero & nonzero  \\
\hline
\end{tabular}
\label{tab:phases}
\end{table}
%%%%%%%%%%%%%%%%%%%%%%%%%%
In this review, we mainly focus on the low-temperature region, with $T\approx 40$~MeV . In this region, the relevant phases are essentially the hadronic phase and the superfluid phase. Within the superfluid phase, there is also a crossover from a strongly coupled BEC regime to a weakly coupled BCS regime. This structure will be discussed below together with the numerical results.

Other possible phases include the hadronic-matter phase, which appears as a narrow region just below the superfluid transition at finite temperature below the pseudo-critical temperature ($T_c$). In this region, $\nq$ is nonzero, while $\qq$ vanishes after the $J\to0$ extrapolation. The region shrinks as the temperature is lowered, supporting the interpretation that it is generated by thermal excitations~\cite{Iida:2019rah,Iida:2024irv}. In addition to these phases, at sufficiently high temperature, chiral symmetry is restored and deconfinement occurs, as in three-colour QCD, leading to the quark--gluon plasma (QGP) phase.

\subsection{Hadronic-superfluid transition}
The mean-field chiral perturbation theory (ChPT) prediction for QC$_2$D is that the superfluid transition occurs when the quark chemical potential reaches one half of the psuedo-scalar mass;
\begin{equation}
  \mu_c\simeq\frac{\mps}{2}.
  \label{eq:onset}
\end{equation}
According to the prediction~\cite{Kogut:2000ek}, the diquark condensate behaves as
\begin{equation}
  \qq = A(\mu-\mu_c)^{1/2},
  \label{eq:ChPT-qq}
\end{equation}
near the transition point. 
Lattice simulations with Wilson and staggered fermions have confirmed this picture at low temperature~\cite{Iida:2019rah,Hands:2006ve,Boz:2019enj,Braguta:2016cpw,Astrakhantsev:2020tdl}.

\begin{figure}[h]
\centering
\includegraphics[width=0.40\textwidth]{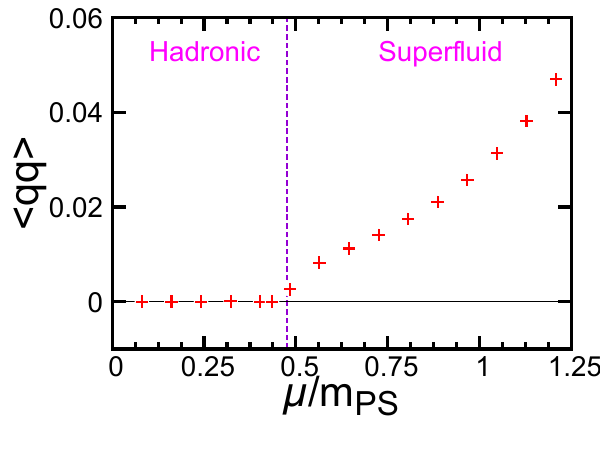}
\includegraphics[width=0.40\textwidth]{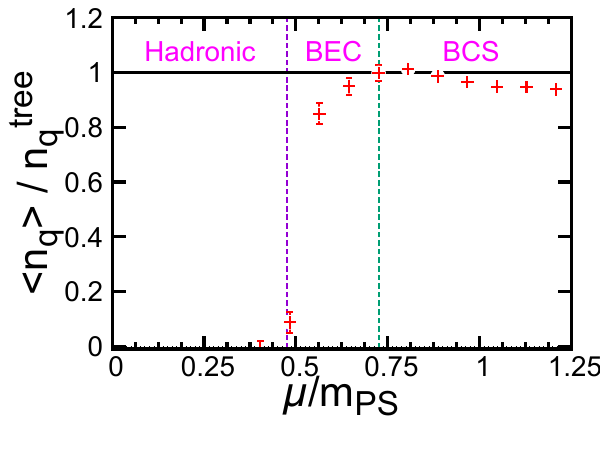}
\caption{
Diquark condensate (Left panel) and quark number density (Right panel) as a function of the quark chemical potential in dense QC$_2$D at $T\simeq40$ MeV.
The data and analysis are taken from Ref.~\cite{Iida:2024irv}.
}
\label{fig:diquark-and-nq}
\end{figure}

In our $T\simeq40$ MeV study~\cite{Iida:2024irv}, we find the value of $\mu_c$ as $\mu_c/\mps\simeq0.48$ by fitting the data using Eq.~\eqref{eq:ChPT-qq}.
The resulting value is close to the expected value as shown in the left panel of Fig.~\ref{fig:diquark-and-nq}. 
%%%%%%%%%%%%%%%%%%%%%%%%%%
%%%%%%%%%%%%%%%%%%%%%%%%%%
The quark number density (the right panel of Fig.~\ref{fig:diquark-and-nq}) also turns on at approximately the same value. 
This provides evidence that a phase transition associated with the breaking of U(1)$_B$ occurs at this point.

\subsection{BEC--BCS crossover}
Inside the superfluid phase, there are at least two distinguished regimes; the BEC and BCS phases.
%%%%%%%%%%%%%%%%%%%%%%%%%%
\begin{figure}[h]
\centering
\includegraphics[width=0.8\textwidth]{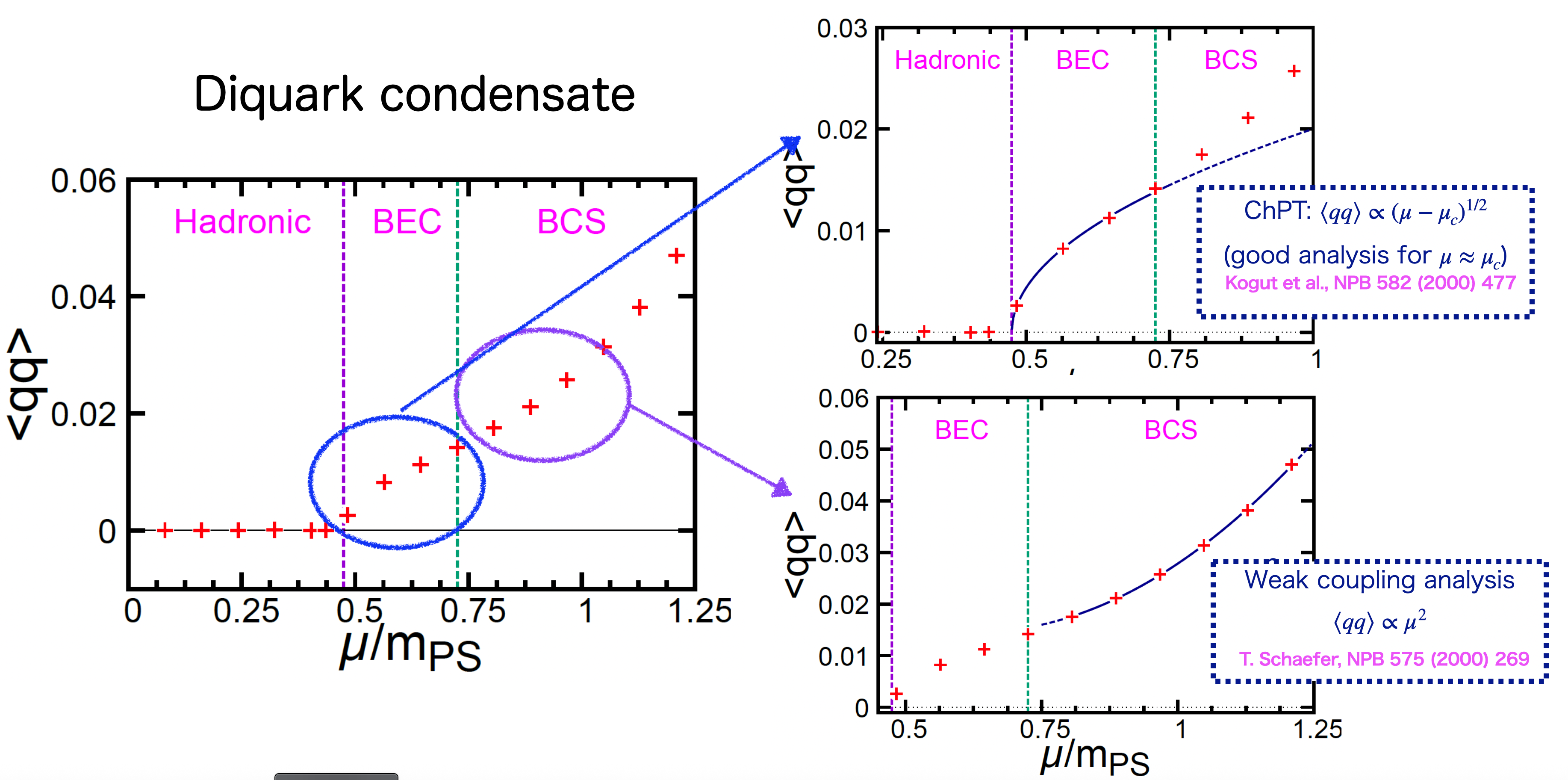}
\caption{
Scaling behaviour of the diquark condensate in dense QC$_2$D at $T\simeq40$ MeV.
In the BEC regime, the data are well described by the ChPT prediction, while in the BCS phase the condensate shows the approximate $\mu^2$ scaling expected by the weak coupling analysis.
The data and analysis are taken from Ref.~\cite{Iida:2024irv}.
}
\label{fig:diquark-scaling}
\end{figure}
%%%%%%%%%%%%%%%%%%%%%%%%%%
Just above onset, tightly bound diquarks condense, and the analytic results from ChPT as an effective theory are expected to be reliable.
At larger chemical potential, the typical distance between quarks decreases, and because of the asymptotic freedom of QCD(-like) theories, the relevant degrees of freedom become quarks near a Fermi surface; the system then becomes BCS-like.
Since the BEC--BCS change is a crossover rather than a symmetry-changing phase transition, its location is definition dependent.

A useful criterion is the ratio of the measured quark number density to the tree-level lattice density, $n_q/n_q^{\rm tree}$~\cite{Hands:2006ve}. Once this ratio approaches unity, the density is well described by weakly interacting quarks and the system is identified as BCS-like. The ratio reaches unity around $\mu/\mps\simeq0.73$ as shown in the right panel of Fig.~\ref{fig:diquark-and-nq}. 

The same region is also consistent with the weak-coupling expectation that the diquark condensate scales roughly as
\begin{equation}
  \qq \propto \mu^2 ,
  \label{eq:bcs}
\end{equation}
reflecting pairing near the Fermi surface \cite{Schafer:1999fe,Hanada:2011ju,Kanazawa:2013crb}. 
On the other hand, in the BEC region the data of $\langle qq \rangle$ can be well fitted by the prediction of ChPT, Eq.~\eqref{eq:ChPT-qq}.
The enlarged plots for each region are shown in Fig.~\ref{fig:diquark-scaling}.
Thus, the guideline for the BEC--BCS crossover determined from $\nq$ also provides a consistent description of the change in the scaling behaviour of $\qq$, from the ChPT-like behaviour in the BEC regime to the weak coupling behaviour expected in the BCS regime.

\subsection{Chiral condensate and gluonic observables}
At high density, chiral symmetry is expected to be restored in both three-colour and two-colour QCD. Correspondingly, the chiral condensate decreases after the onset of superfluidity in several QC$_2$D simulations. 
In the BEC regime, the data in Ref.~\cite{Iida:2024irv} follow the ChPT predictions,
\begin{equation}
\qbarq(\mu)\propto\frac{1}{\mu^2},
  \label{eq:chiral}
\end{equation}
while at higher density it seems to become flatter at nonzero values. 
Because Wilson fermions explicitly break chiral symmetry, this behaviour should be interpreted carefully. 
Nevertheless, it indicates a tendency toward chiral-symmetry restoration. 
Recent studies have also shown that correlation functions of chiral partners become degenerate in the high-density region~\cite{Iida:2026dmw}.
This provides further support for the tendency toward chiral restoration at high density.

The gluonic sector shows a different and quite interesting behaviour. Early studies discussed the possibility that the dense BCS regime might deconfine, but more recent work suggests that confinement can persist at low temperature even at large chemical potential. In a work of the static quark-antiquark potential, the potential remains linearly rising in the explored dense regime, and the string tension stays nonzero \cite{Ishiguro:2021yxr}. Independent Wilson and staggered studies also indicate that the deconfinement temperature remains of order $100$ MeV at high density \cite{Boz:2019enj,Begun:2022bxj}.

The topological susceptibility provides another view of the same issue. At high temperature, where the Polyakov loop grows, topological fluctuations are suppressed as the chemical potential is increased. At low temperature, however, the simulations at $T\simeq80$ MeV and $T\simeq40$ MeV find an almost $\mu$-independent topological susceptibility~\cite{Iida:2019rah,Iida:2024irv}. 
This suggests that the cold dense regime retains nonperturbative gluonic features.

This separation between quark and gluon diagnostics is one of the central lessons of the present lattice results. 
In high-temperature QCD the chiral condensate, topological susceptibility and Polyakov loop all change rapidly around the pseudo-critical temperature, giving a relatively coherent picture of the transition from hadrons to the QGP. 
In cold dense QC$_2$D, by contrast, the quark number density and diquark condensate indicate a weakly interacting quark picture, but the Polyakov loop, static potential and topology do not show the same kind of deconfining response.
Thus, the BCS-like dense regime is not simply the low-temperature continuation of the high-temperature plasma.

%%%%%%%%%%%%%%%%%%%%%%%%%%%%%%%%%%%%%%%%%%%%%%%%%%%%%%%%%
\section{EoS and sound velocity}\label{sec:eos}
%%%%%%%%%%%%%%%%%%%%%%%%%%%%%%%%%%%%%%%%%%%%%%%%%%%%%%%%%
\subsection{Thermodynamic quantities}
We now turn to the EoS and the sound velocity in the low-temperature regime discussed above, namely around $T\simeq 40$ MeV.
The pressure at finite density is obtained from the quark number density by 
\begin{equation}
  p(\mu)-p(\mu_c)=\int_{\mu_c}^{\mu} n_q(\mu')\,d\mu' ,
  \label{eq:pressure}
\end{equation}
in the thermodynamic limit. Here, $p(\mu)=0$ in $\mu \leq \mu_c$.
The trace anomaly is evaluated from derivatives of the lattice action with respect to the lattice spacing, which is explicitly given by 
\begin{equation}
  e-3p =
  -\frac{T}{V}
  \left[
    a\frac{\partial\beta}{\partial a}
      \left\langle\frac{\partial S}{\partial\beta}\right\rangle
   +a\frac{\partial\kappa}{\partial a}
      \left\langle\frac{\partial S}{\partial\kappa}\right\rangle
   +a\frac{\partial j}{\partial a}
      \left\langle\frac{\partial S}{\partial j}\right\rangle
  \right]_{\rm sub},
  \label{eq:trace}
\end{equation}
where the subtraction removes the vacuum contribution. In present analyses, the first two terms are evaluated using the nonperturbative beta functions from the scale-setting study, while the last term is assumed to vanish in the $j\to0$ limit. This assumption is plausible but should be checked more carefully in future work.
Combining the pressure and the trace anomaly gives the internal energy density as a function of $\mu$.

\begin{figure}[h]
\centering
\includegraphics[width=0.6\textwidth]{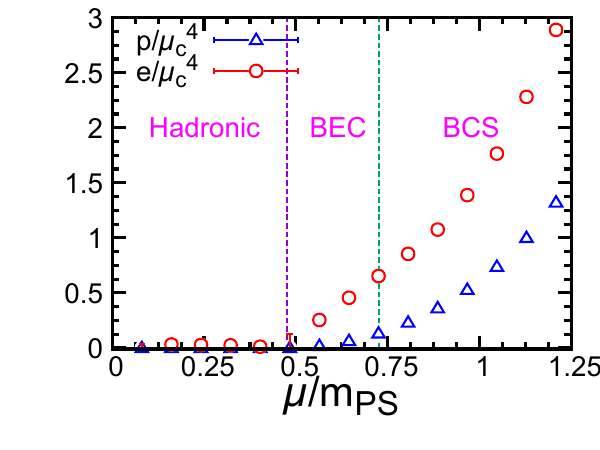}
\caption{Pressure and internal energy density as functions of the quark chemical potential in dense QC$_2$D. The plot is originally obtained from the $T\simeq40$ MeV lattice study of Ref.~\cite{Iida:2024irv}.}
\label{fig:eos}
\end{figure}

The resulting pressure and internal energy are shown in Fig.~\ref{fig:eos}. 
%%%%%%%%%%%%%%%%%%%%%%%%%%%
%%%%%%%%%%%%%%%%%%%%%%%%%%%
These quantities grow rapidly after the onset of superfluidity, while both $p$ and $e$ remain consistent with zero in the hadronic phase.

\subsection{Sound velocity and conformal-bound violation}
The sound velocity is then extracted from
\begin{equation}
  c_{\rm s}^2/c^2 =
  \left.\frac{\partial p}{\partial e}\right|_{s}
  \simeq
  \left.\frac{\partial p/\partial\mu}{\partial e/\partial\mu}\right|_T ,
  \label{eq:sound}
\end{equation}
where the final expression is appropriate when thermal effects are small. 
The comparison of $T\simeq80$ MeV and $T\simeq40$ MeV data indicates that this approximation is reliable in the cold regime studied so far \cite{Iida:2022hyy,Iida:2024irv}.
The ChPT~\cite{Son_2001,Kogut:2000ek,Hands:2006ve} predicts
\begin{equation}
  (c_{\rm s}^2/c^2)_{\rm ChPT}=
  \frac{1-\mu_c^4/\mu^4}{1+3\mu_c^4/\mu^4}.
  \label{eq:chptsound}
\end{equation}

%%%%%%%%%%%%%%%%%%%%%%%%%%%
\begin{figure}[h]
\centering
\includegraphics[width=0.6\textwidth]{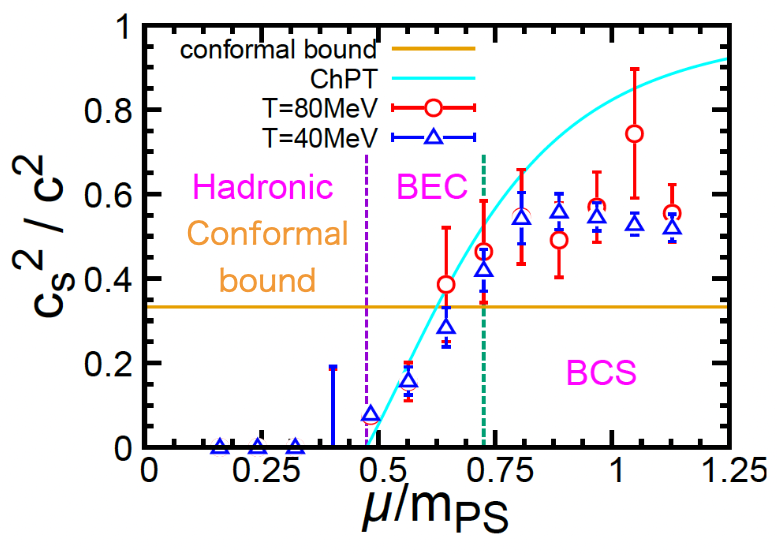}
\caption{Squared sound velocity in dense QC$_2$D. The $T\simeq80$ MeV result was first reported in Ref.~\cite{Iida:2022hyy}, and the combined comparison with the $T\simeq40$ MeV data was obtained in Ref.~\cite{Iida:2024irv}. The horizontal orange line marks the conformal value $c_{\rm s}^2/c^2=1/3$.}
\label{fig:sound}
\end{figure}
%%%%%%%%%%%%%%%%%%%%%%%%%%%
Figure~\ref{fig:sound} shows the sound velocity obtained from the simulations at $T\simeq40$ MeV and $T\simeq80$ MeV. 
The results at $T\simeq40$ MeV and $T\simeq80$ MeV show no difference within errors.
In the hadronic phase, the sound velocity is consistent with zero, and it becomes nonzero after the transition to the superfluid phase. As in the cases of the diquark condensate and the chiral condensate, the data in the BEC regime agree well with the ChPT prediction, Eq.~\eqref{eq:chptsound}, shown by the cyan curve. Once the onset chemical potential $\mu_c$ is fixed, this ChPT prediction contains no free parameters, and the agreement provides a nontrivial check of the lattice calculation.

On the other hand, the ChPT expression gives $c_{\rm s}^2/c^2\to1$ in the high-density limit, which is unphysical. Thus, the ChPT result should be regarded as applicable only near the onset of the superfluid phase. In the BCS-like regime, the lattice data deviate from the ChPT prediction and do not continue to increase as a function of $\mu$. The important observation is that the sound velocity nevertheless exceeds the conformal value, $c_{\rm s}^2/c^2=1/3$, in the BCS phase~\cite{Iida:2022hyy,Iida:2024irv}. Such behaviour had not been clearly observed in previous lattice calculations of QCD-like theories.

%%%%%%%%%%%%%%%%%%%%%%%%%%%%%%%%%%%%%%%%%%%%%%%%%%%%%%%%%
\section{Conclusions and outlook}\label{sec:summary}
%%%%%%%%%%%%%%%%%%%%%%%%%%%%%%%%%%%%%%%%%%%%%%%%%%%%%%%%%
Dense two-colour QCD has developed into a quantitative laboratory for cold dense strong-interaction matter. The phase diagram is now rather coherent across independent simulations. The hadronic-superfluid transition occurs around $\mu_c\simeq\mps/2$, the BEC--BCS crossover is visible through the quark number density and the diquark condensate.
Furthermore, the low-temperature dense regime appears to retain confinement and unsuppressed topological fluctuations over the range explored so far.

The EoS gives the most provocative lesson. In the cold superfluid regime, lattice data show that the squared sound velocity can exceed the conformal value $1/3$. This result does not by itself solve the EoS of neutron-star matter, but it gives a first-principles demonstration that QCD-like dense matter can be stiff. It also sharpens the question of which features are universal among dense gauge theories and which rely on the special colour-singlet nature of the QC$_2$D\ diquark condensate.

Recent QC$_2$D\ work by other groups also reports a large sound velocity in the high-density regime \cite{Hands:2024nkx}. 
Another sign-problem-free theory to dense matter is three-colour QCD at finite isospin chemical potential. 
In that case the chemical potentials of the up and down quarks have opposite signs, and the product of the two fermion determinants is real and positive. The onset to the superfluid phase can be controlled by charged pion condensation rather than by diquark condensation. 
An external pionic source is introduced to stabilize the numerical simulations in high-density regime, and observables are extrapolated to zero source.
In such a QCD-like theory, the EoS and sound velocity have been studied without the baryon sign problem, and have shown the violations of the conformal value~\cite{Brandt:2022hwy,Abbott:2023coj,Abbott:2024vhj}

Real neutron-star matter is governed by three-colour QCD at baryon chemical potential, not by QC$_2$D. Therefore the lattice results reviewed here should not be inserted directly into astrophysical equations of state.
Their value is instead conceptual: they provide a first-principles example of strongly interacting dense matter with a large sound velocity.
This point is relevant because many phenomenological analyses of neutron-star data favour a stiffening of matter at several times nuclear saturation density \cite{Baym:2017whm,Altiparmak:2022bke,Brandes:2022nxa,Annala:2023cwx}. Effective descriptions, including quarkyonic models, Nambu--Jona-Lasinio-type models, functional renormalization group studies and holographic constructions, have also found peaks in the sound velocity \cite{McLerran:2018hbz,Kojo:2021ugu,Braun:2022jme,Fukushima:2024gmp}.

The next steps are clear. Continuum and thermodynamic extrapolations are needed, especially for the sound velocity and gluonic observables. The chiral limit and lighter quark masses should be explored. More detailed comparisons among Wilson, staggered and improved actions will be essential. Finally, the relation between QC$_2$D, isospin-dense QCD and phenomenological dense QCD should be developed in a common language. These efforts will make sign-problem-free theories more useful as controlled laboratories for the physics of compact stars.

\ack{The author thanks K. Iida, K. Ishiguro, T.-G. Lee, K. Murakami and D. Suenaga for collaborations on the original works reviewed here. The author also thanks S. Hands, T. Hatsuda, T. Kojo, J.-I. Skullerud, N. Yamamoto  for valuable discussions. }

\funding{The author is supported by JSPS KAKENHI Grant Nos.~25K01001 and 23H05439, % Kiban-B, Kiban-S
JST SQAI Grant No.~JPMJPF2221, % SQAI
and JST CREST Grant No.~JPMJCR24I3. % CREST
This work is supported by the Program for Promoting Researches on the Supercomputer ``Fugaku'' (Simulation for basic science: from fundamental laws of particles to creation of nuclei; Simulation for basic science: approaching the new quantum era), and the Joint Institute for Computational Fundamental Science (JICFuS), Grant No.~JPMXP1020230411. % Fugaku
This work is supported by the Center for Gravitational Physics and Quantum Information (CGPQI) at the Yukawa Institute for Theoretical Physics.}

\data{No new data were generated for this review.}

\bibliographystyle{unsrt}
\bibliography{2color}

\end{document}